\documentclass[11pt,a4paper]{article}
\pdfoutput=1
\usepackage{jheppub}
\usepackage{tikz}
\usepackage{simpler-wick}
\usepackage{bbm}

\DeclareMathOperator{\Tr}{Tr}

\newcommand{\ri}{\mathrm{i}}
\renewcommand{\th}{\theta}

\newcommand{\cob}{\delta}

\newcommand{\hf}{\frac{1}{2}}

\newcommand{\til}[1]{\widetilde{#1}}

\newcommand{\lap}{\Delta}
\newcommand{\bra}{\langle}
\newcommand{\ket}{\rangle}
\newcommand{\la}{\lambda}

\newcommand{\h}[1]{\widehat{#1}}
\newcommand{\bt}{\beta}

\newcommand{\al}{\alpha}

\newcommand{\rt}[1]{\sqrt{#1}}
\newcommand{\cO}{\mathcal{O}}

\newcommand{\cH}{\mathcal{H}}

\newcommand{\cB}{\mathcal{B}}

\newcommand{\id}{\mathbbm{1}}

\makeatletter
\gdef\@fpheader{}
\makeatother
\begin{document}
\title{Baby universe operators in the ETH matrix model of double-scaled SYK}

\author{Kazumi Okuyama}

\affiliation{Department of Physics, 
Shinshu University, 3-1-1 Asahi, Matsumoto 390-8621, Japan}

\emailAdd{kazumi@azusa.shinshu-u.ac.jp}

\abstract{
We consider the baby universe operator $\mathcal{B}_a$
in the double-scaled SYK (DSSYK) model, which creates a baby universe of size $a$.
We find that $\mathcal{B}_a$ is written in terms of the transfer matrix
$T$, and vice versa.
In particular, the identity operator on the chord Hilbert space
is expanded as a linear combination
of $\mathcal{B}_a$, which implies that 
the disk partition function of DSSYK is written as a 
linear combination of trumpets.
We also find that the thermofield double state of DSSYK
is generated by a pair of baby universe operators, which corresponds to a
double trumpet.
This can be thought of as a concrete realization of the idea of 
ER=EPR.
}

\maketitle

\section{Introduction}\label{sec:intro}

Baby universes have a long history in the study of quantum gravity.
In \cite{Coleman:1988cy}, Coleman famously argued that the effect
of baby universes leads to a loss of quantum coherence
and the presence of the so-called $\al$-parameter.
In the modern language, the integral over the $\al$-parameter
is interpreted as an ensemble average in the boundary theory which is 
holographically dual to the bulk quantum gravity \cite{Marolf:2020xie}.
In particular, the duality between Sachdev-Ye-Kitaev (SYK) model \cite{Sachdev1993,Kitaev1,Kitaev2}
and Jakiew-Teitelboim (JT) gravity \cite{Jackiw:1984je,Teitelboim:1983ux}
has been extensively studied in the literature as an example of holography
involving ensemble averaging.\footnote{
It is conjectured in \cite{McNamara:2020uza}
that the baby universe Hilbert space of a unitary theory of quantum gravity
in $d>3$
is one-dimensional and hence there is no $\al$-parameter.
This conjecture does not apply to JT gravity in $d=2$.}

In this paper,
we consider the so-called double-scaled SYK (DSSYK)
model \cite{Cotler:2016fpe,Berkooz:2018jqr}
as a useful toy model to study the bulk Hilbert space of quantum gravity.
As shown in \cite{Berkooz:2018jqr}, 
the computation of the correlators in DSSYK boils down to a counting
problem of chord diagrams. This counting problem 
can be exactly solved by 
introducing the transfer matrix $T$
written in terms of the $q$-deformed oscillator $A_\pm$, 
which creates or annihilates the chords.
In the absence of matter particles, 
it is argued in \cite{Lin:2022rbf} that the bulk Hilbert space of
DSSYK is identified with the Fock space of 
$q$-deformed oscillator, called the chord Hilbert space.

It turns out that we can define the baby universe operator
$\cB_a$ in DSSYK which creates a baby universe of size $a$,
in a similar manner as the one in JT gravity considered in 
\cite{Penington:2023dql}.
We find that the baby universe operator
$\cB_a$ can be expanded as a combination of the transfer matrix $T$, 
and vice versa.
We also find that the identity operator $\id$ on the chord Hilbert space
is written as a linear combination of the baby universe operators.
This implies that the disk partition function of 
DSSYK can be written as a linear combination of ``trumpets''.
We also consider the thermofield double state in DSSYK.
We find that it is generated by a pair of baby universe operators
corresponding to the ``double trumpet'' \cite{Saad:2018bqo,Saad:2019lba}.
This can be thought of as a concrete realization of the idea of 
``$\text{ER}=\text{EPR}$'' \cite{VanRaamsdonk:2010pw,Maldacena:2013xja}.

This paper is organized as follows. 
In section \ref{sec:review}, we briefly review 
the known results of DSSYK in \cite{Berkooz:2018jqr}.
In section \ref{sec:baby}, we study the properties
of the baby universe operator in DSSYK.
We show that the identity operator $\id$ on the chord Hilbert space
and the transfer matrix $T$ can be written as a linear combination
of baby universe operators. 
In section \ref{sec:TFD}, we consider the thermofield double state
in DSSYK and find that it is written in terms of a pair of baby universe operators.
In section \ref{sec:JT}, we discuss the similarity and differences
of the baby universe operators in JT gravity and DSSYK.
Finally we conclude in section \ref{sec:conclusion}
with some discussion on the future problems.

\section{Review of double-scaled SYK}\label{sec:review}

In this section we will briefly review the 
result of DSSYK 
in \cite{Berkooz:2018jqr}.
SYK model is defined by the Hamiltonian for 
$N$ Majorana fermions $\psi_i~(i=1,\cdots,N)$
obeying $\{\psi_i,\psi_j\}=2\cob_{i,j}$
with all-to-all $p$-body interaction
\begin{equation}
\begin{aligned}
 H=\ri^{p/2}\sum_{1\leq i_1<\cdots<i_p\leq N}
J_{i_1\cdots i_p}\psi_{i_1}\cdots\psi_{i_p},
\end{aligned} 
\label{eq:SYK}
\end{equation}
where $J_{i_1\cdots i_p}$ is a random coupling drawn from the Gaussian distribution.
DSSYK is defined by the scaling limit
\begin{equation}
\begin{aligned}
 N,p\to\infty\quad\text{with}\quad \la=\frac{2p^2}{N}:\text{fixed}.
\end{aligned} 
\label{eq:scaling}
\end{equation}
The average of the moment $\Tr H^k$ 
over the random coupling $J_{i_1\cdots i_p}$
reduces to a counting problem
of the intersection number of chord diagrams. 
As shown in \cite{Berkooz:2018jqr}, this problem can be
solved by introducing the transfer matrix $T$
\begin{equation}
\begin{aligned}
 \bra \Tr H^k\ket_J=\sum_{\text{chord diagrams}}q^{\#(\text{intersections})}
=\bra 0|T^k|0\ket
\end{aligned}
\label{eq:moment} 
\end{equation}
where $q=e^{-\la}$
and $T$ is given by the 
$q$-deformed oscillator $A_{\pm}$\footnote{Our definition of $T$
differs from that in \cite{Berkooz:2018jqr} by an overall minus sign.
This minus sign does not change the result of \eqref{eq:moment}
since the moment $\bra \Tr H^k\ket_J$ vanishes for odd $k$. 
}
\begin{equation}
\begin{aligned}
 T=-E_0X,\quad X=\frac{A_{+}+A_{-}}{2},\quad E_0=\frac{2}{\rt{1-q}}.
\end{aligned}
\label{eq:TX-def} 
\end{equation}
Note that $A_{\pm}$ obey the commutation relation
\begin{equation}
\begin{aligned}
 A_{-}A_{+}-qA_{+} A_{-}=1-q,
\end{aligned} 
\end{equation}
and they act on the chord number state $|n\ket$ as
\begin{equation}
\begin{aligned}
 A_{+}|n\ket=\rt{1-q^{n+1}}|n+1\ket,\quad
A_{-}|n\ket=\rt{1-q^{n}}|n-1\ket.
\end{aligned} 
\label{eq:Apm-n}
\end{equation}
Here $|n\ket$ denotes
the state with $n$ chords 
\begin{equation}
\begin{aligned}
 \begin{tikzpicture}[scale=0.75]
\draw (0,-1)--(0,1);
\draw (0.5,-1)--(0.5,1);
\draw (1,-1)--(1,1);
\draw (1.5,-1)--(1.5,1);
\draw (2,-1)--(2,1);
\draw (2.5,-1)--(2.5,1);
\draw (3,-1)--(3,1);
\draw[dashed] (-0.5,0)--(3.5,0);
\draw (-1,0) node [left]{$|n\ket=$};
\draw (1.5,1) node [above] {$\overbrace{~\hskip23mm~}^{}$};
\draw (1.5,1.5) node [above] {$n~\text{chords}$};
\end{tikzpicture}
\end{aligned} 
\label{eq:def-n}
\end{equation}
and the dashed line in \eqref{eq:def-n} represents
a constant (Euclidean) time slice.
The Hilbert space $\cH$ spanned by the chord number states
is called the chord Hilbert space
\begin{equation}
\begin{aligned}
 \cH=\bigoplus_{n=0}^\infty\mathbb{C}|n\ket.
\end{aligned} 
\label{eq:H-chord}
\end{equation}
In \cite{Lin:2022rbf}, the chord Hilbert space is identified as the
Hilbert space of bulk quantum gravity and
the chord number $n$ represents the bulk geodesic length.

The operator $X$ in \eqref{eq:TX-def} can be diagonalized 
by the $|\th\ket$-basis and the eigenvalue is given by
$\cos\th$
\begin{equation}
\begin{aligned}
X|\th\ket=\cos\th|\th\ket.
\end{aligned} 
\label{eq:eigen-th}
\end{equation}
From \eqref{eq:TX-def}, one can see that the eigenvalue 
$E(\th)$ of the transfer matrix
$T$ is given by
\begin{equation}
\begin{aligned}
 T|\th\ket=E(\th)|\th\ket,\quad E(\th)=-E_0\cos\th,
\end{aligned} 
\label{eq:E-th}
\end{equation} 
which is bounded from above and below
\begin{equation}
\begin{aligned}
 -E_0\leq E(\th)\leq E_0.
\end{aligned} 
\end{equation}

The overlap of $|\th\ket$ and $|n\ket$  is
given by the $q$-Hermite polynomial $H_n(\cos\th|q)$
\begin{equation}
\begin{aligned}
 \bra \th|n\ket=\bra n|\th\ket=\frac{H_{n}(\cos\th|q)}{\rt{(q;q)_n}},
\end{aligned} 
\label{eq:n-th}
\end{equation}
where $(q;q)_n$ denotes the $q$-Pochhammer symbol
\begin{equation}
\begin{aligned}
 (a;q)_n=\prod_{k=0}^{n-1}(1-aq^k),
\end{aligned} 
\end{equation}
and the $q$-Hermite polynomial is defined by
\begin{equation}
\begin{aligned}
 H_n(\cos\th|q)=\sum_{k=0}^n\frac{(q;q)_n}{(q;q)_k(q;q)_{n-k}}e^{\ri(n-2k)\th}.
\end{aligned} 
\label{eq:q-H}
\end{equation}
The $q$-Hermite polynomials are orthogonal with respect to the measure factor
\footnote{
Throughout this paper, we use the standard notation 
\begin{equation}
\begin{aligned}
 (a_1,\cdots,a_m;q)_\infty=\prod_{j=1}^m(a_j;q)_\infty,\quad
(ae^{\pm\ri\th};q)_\infty=(ae^{\ri\th},ae^{-\ri\th};q)_\infty.
\end{aligned} 
\end{equation}
}
\begin{equation}
\begin{aligned}
 \mu(\th)=(q,e^{\pm2\ri\th};q)_\infty.
\end{aligned} 
\label{eq:mu-def}
\end{equation}
and the orthogonality relation is written as
\begin{equation}
\begin{aligned}
  \bra n|m\ket=\int_0^\pi\frac{d\th}{2\pi}\mu(\th)\bra n|\th\ket\bra \th|m\ket
=\cob_{n,m}.
\end{aligned} 
\label{eq:cob-nm}
\end{equation}
One can also show that
\begin{equation}
\begin{aligned}
 \bra\th|\th'\ket=\frac{2\pi}{\mu(\th)}\cob(\th-\th'),
\end{aligned} 
\label{eq:th-inner}
\end{equation}
and the resolution of the identity is given by
\begin{equation}
\begin{aligned}
 \id=\sum_{n=0}^\infty|n\ket\bra n|=\int_0^\pi\frac{d\th}{2\pi}\mu(\th)
|\th\ket\bra\th|.
\end{aligned} 
\label{eq:def-id}
\end{equation} 

As discussed in \cite{Berkooz:2018jqr}, we can 
also introduce the matter operator 
$\cO_\lap$ in DSSYK
\begin{equation}
\begin{aligned}
 \cO_\lap=\ri^{s/2}\sum_{1\leq i_1<\cdots< i_s\leq N}K_{i_1\cdots i_s}\psi_{i_1}\cdots\psi_{i_s}
\end{aligned} 
\label{eq:matter-M}
\end{equation}
with a Gaussian random coefficient $K_{i_1\cdots i_s}$
which is drawn independently from the random coupling
$J_{i_1\cdots i_p}$ in the Hamiltonian \eqref{eq:SYK}. The number $s$ of fermions in
\eqref{eq:matter-M} is related to the dimension $\lap$ of the operator
$\cO_\lap$ by
\begin{equation}
\begin{aligned}
 \lap=\frac{s}{p}.
\end{aligned} 
\end{equation} 
Again, the correlator of $\cO_\lap$'s
is written as a  
counting problem of the chord diagrams
\begin{equation}
\begin{aligned}
 \sum_{\text{chord diagrams}}q^{\#(H\text{-}H\,\text{intersections})}
q^{\lap_i\#(H\text{-}\cO_{\lap_i}\,\text{intersections})}
q^{\lap_i\lap_j\#(\cO_{\lap_i}\text{-}\cO_{\lap_j}\,\text{intersections})}.
\end{aligned} 
\label{eq:chord-count}
\end{equation}
Note that there appear two types of chords in this computation: 
the $H$-chord and 
the matter chord coming from the Wick contraction of random couplings 
$J_{i_1\cdots i_p}$ and $K_{i_1\cdots i_s}$, respectively.

\subsection{Higher genus corrections and ETH matrix model}
In order to study the baby universe operators in DSSYK,
it is useful to consider the ETH\footnote{
ETH stands for the ``eigenvalue thermalization hypothesis''.} matrix model of DSSYK introduced in \cite{Jafferis:2022wez}
\begin{equation}
\begin{aligned}
 Z=\int_{L\times L}dH e^{-L\Tr V(H)},
\end{aligned} 
\label{eq:ETH}
\end{equation}
where $H$ is a $L\times L$ random hermitian matrix
and 
$L$ is given the dimension of the Hilbert space of $N$ Majorana fermions
of the SYK model
\begin{equation}
\begin{aligned}
 L=2^{N/2}.
\end{aligned} 
\end{equation} 
In \eqref{eq:ETH}, we have ignored the effect of matter operators for simplicity;
if we include the effect of matter loops the ETH matrix model is described by 
a two-matrix model \cite{Jafferis:2022wez}.
The potential $V(H)$ in \eqref{eq:ETH} is determined by
 matching the eigenvalue density at the disk level
\begin{equation}
\begin{aligned}
 \mu(\th)d\th=\rho_0(E)dE,\qquad (E=-E_0\cos\th),
\end{aligned} 
\end{equation}
where $\mu(\th)$ is the measure factor of DSSYK in \eqref{eq:mu-def}
while $\rho_0(E)$ is the genus-zero eigenvalue density of
the ETH matrix model \eqref{eq:ETH}.

By construction, the ETH matrix model agrees with DSSYK 
at the disk level. However, 
the ETH matrix model does not necessarily agrees with DSSYK 
beyond the disk level
 \cite{Jafferis:2022wez}
(see also \cite{Berkooz:2020fvm}).
Nevertheless, it is interesting to 
study the large $L$ expansion of the 
ETH matrix model \eqref{eq:ETH} in its own right.
Indeed, as shown in  \cite{Okuyama:2023kdo},
the higher genus corrections of the ETH matrix model \eqref{eq:ETH}
have a similar structure as the JT gravity matrix model \cite{Saad:2019lba}.
In the ETH matrix model,
one can define the trumpet and the discrete analogue 
of the Weil-Petersson volume 
and the
higher genus corrections
are given by some combination of them \cite{Okuyama:2023kdo}
in a completely parallel way as the JT gravity case. 
In the rest of this paper, by ``trumpet'' and ``wormhole'' of DSSYK we mean
those in the ETH matrix model of DSSYK,
in the sense that we consider the large $L$ expansion of  
\eqref{eq:ETH} in its own right.

\section{Baby universe operators}\label{sec:baby}
In this section, we introduce the baby universe operator in DSSYK and
study its properties.
As discussed in \cite{Jafferis:2022wez,Okuyama:2023byh},
we can define the ``trumpet'' in DSSYK 
in a similar manner as JT gravity
\begin{equation}
\begin{tikzpicture}[scale=0.75]
\draw (0,2) arc [start angle=90,end angle=270, x radius=1, y radius=2];
\draw[dashed] (0,2) arc [start angle=90,end angle=-90, x radius=1, y radius=2];
\draw[thick,red] (5.25,0) arc [start angle=360,end angle=0, x radius=0.25, y radius=0.5];
\draw (0.14,1.99) to [out=-32,in=180] (5,0.5);
\draw (0.03,-2.02) to [out=31,in=180] (5,-0.5);
\draw (0,-2.1) node [below]{$\bt$};
\draw (5,-0.5) node [below]{$a$};
\draw (3.5,0) node [left]{$\text{trumpet}$};
\draw (5.8,0) node [right]{$\displaystyle =~~
I_a(\bt E_0)$,};
\end{tikzpicture}
\label{eq:trumpet-fig}
\end{equation}
where $I_\nu(z)$ denotes the modified Bessel function
of the first kind, 
and $\bt$ and $a$ represent
the length of the asymptotic boundary and the length of the geodesic loop,
respectively.
One important difference from JT gravity is that
the length $a$ of the geodesic loop 
becomes discrete in DSSYK
\begin{equation}
\begin{aligned}
 a\in\mathbb{Z}_{\geq0}.
\end{aligned} 
\end{equation}
Using the integral representation of the modified Bessel function,
the trumpet partition function in \eqref{eq:trumpet-fig} is written as
\begin{equation}
\begin{aligned}
 Z_{\text{trumpet}}(\bt,a)=I_a(\bt E_0)
=\int_0^\pi\frac{d\th}{2\pi}e^{-\bt E(\th)}2\cos(a\th),
\end{aligned} 
\label{eq:trumpet-int}
\end{equation}
where $E(\th)$ is the eigenvalue of the transfer matrix $T$
in \eqref{eq:E-th}.

As argued in \cite{Saad:2019pqd}, $Z_{\text{trumpet}}(\bt,a)$
can be interpreted as the emission amplitude of a baby universe of size $a$.
We can introduce the baby universe operator $\cB_a$
in such a way that
$Z_{\text{trumpet}}(\bt,a)$ is written as
\begin{equation}
\begin{aligned}
 Z_{\text{trumpet}}(\bt,a)=\bra 0|e^{-\bt T} \cB_a|0\ket,
\end{aligned} 
\end{equation}
where $|0\ket$ is the $0$-chord state.
Using $\bra\th|0\ket=1$,
one can easily see that the following expression of 
$\cB_a$ reproduces
the result of $Z_{\text{trumpet}}(\bt,a)$ in \eqref{eq:trumpet-int}
\begin{equation}
\begin{aligned}
 \cB_a=\int_0^\pi\frac{d\th}{2\pi}\,2\cos(a\th)|\th\ket\bra\th|.
\end{aligned} 
\label{eq:Ba-def}
\end{equation}
From \eqref{eq:th-inner}, we can see that
$\cB_a$ in \eqref{eq:Ba-def} is diagonal in the $|\th\ket$-basis
\begin{equation}
\begin{aligned}
 \cB_a|\th\ket=\frac{2\cos(a\th)}{\mu(\th)}|\th\ket.
\end{aligned} 
\end{equation}
This is an analogue of the relation obeyed by
the baby universe operator
in JT gravity (see eq.(4.3) in \cite{Penington:2023dql}).
As discussed in \cite{Penington:2023dql}, 
the baby universe operator $\cB_a$ creates a hole of size $a$.
Since $X$ in \eqref{eq:TX-def} and $\cB_a$ are both
diagonal in the $|\th\ket$-basis,
they commute with each other
\begin{equation}
\begin{aligned}
 {[}X,\cB_a{]}=0.
\end{aligned} 
\end{equation}

In the rest of this section, we will show that $\cB_a$ is written in terms
of $X$, and vice versa.
Using the relation found in \cite{Okuyama:2023byh}
\begin{equation}
\begin{aligned}
 \frac{2\cos(a\th)}{\mu(\th)}=
\sum_{k=0}^\infty\frac{H_{a+2k}(\cos\th|q)}{(q;q)_k(q;q)_{a+k}},
\end{aligned} 
\label{eq:cos/mu}
\end{equation}
we find
\begin{equation}
\begin{aligned}
 \cB_a&=\int_0^\pi\frac{d\th}{2\pi}\mu(\th)\frac{2\cos(a\th)}{\mu(\th)}|\th\ket\bra\th|\\
&=\int_0^\pi\frac{d\th}{2\pi}\mu(\th)
\sum_{k=0}^\infty\frac{H_{a+2k}(\cos\th|q)}{(q;q)_k(q;q)_{a+k}}
|\th\ket\bra\th|\\
&=\sum_{k=0}^\infty\frac{H_{a+2k}(X|q)}{(q;q)_k(q;q)_{a+k}}.
\end{aligned} 
\label{eq:Ba-X}
\end{equation}
In the last equality we have used the relation valid for arbitrary function
$f(X)$ of $X$
\begin{equation}
\begin{aligned}
 f(X)=\int_0^\pi\frac{d\th}{2\pi}\mu(\th)f(\cos\th)|\th\ket\bra\th|.
\end{aligned} 
\label{eq:f(X)}
\end{equation} 
\eqref{eq:Ba-X} is 
the desired expansion of the baby universe operator $\cB_a$
in terms of $X$.

Conversely, it turns out that various operators $f(X)$ on the chord Hilbert space
$\cH$ can be expanded as a linear combination of $\cB_a$.
As an example, let us consider the identity operator $\id$ in \eqref{eq:def-id}.
Using the Jacobi triple product identity, we can easily show that
$\mu(\th)$ in \eqref{eq:mu-def} is expanded as
\begin{equation}
\begin{aligned}
 \mu(\th)=\sum_{r\in\mathbb{Z}}
(-1)^rq^{\hf r(r+1)}2\cos(2r\th).
\end{aligned} 
\label{eq:Jacobi}
\end{equation}
Plugging this into \eqref{eq:def-id}
and using \eqref{eq:Ba-def},
we find that the identity operator $\id$ 
is written as
\begin{equation}
\begin{aligned}
 \id=
\sum_{r\in\mathbb{Z}}
(-1)^rq^{\hf r(r+1)}\cB_{2r}.
\end{aligned} 
\label{eq:id-B}
\end{equation}
Here and in what follows, $\cB_{a}$ should be understood as $\cB_{|a|}$, i.e., 
the baby universe operator $\cB_a$ depends only on the absolute value of $a$
\begin{equation}
\begin{aligned}
 \cB_a=\cB_{-a}=\cB_{|a|}.
\end{aligned} 
\end{equation}
As an application of our result \eqref{eq:id-B}, let us consider the disk partition
function $\bra 0|e^{-\bt T}|0\ket$.
Inserting the identity operator $\id$ in \eqref{eq:id-B}, we find
\begin{equation}
\begin{aligned}
 \bra 0|e^{-\bt T}|0\ket&=\bra 0|\id \cdot e^{-\bt T}|0\ket\\
&=\sum_{r\in\mathbb{Z}}
(-1)^rq^{\hf r(r+1)}\bra 0|\cB_{2r}  e^{-\bt T}|0\ket\\
&=\sum_{r\in\mathbb{Z}}
(-1)^rq^{\hf r(r+1)}I_{2r}(\bt E_0).
\end{aligned} 
\label{eq:disk}
\end{equation}
This reproduces the result of disk partition function in \cite{Berkooz:2018jqr}.
Our result \eqref{eq:disk} implies that the disk partition function
is written as a linear combination of the trumpet
in \eqref{eq:trumpet-fig}, which
is schematically depicted as
\begin{equation}
\begin{tikzpicture}[scale=0.75]
\draw (0,2) arc [start angle=90,end angle=270, x radius=1, y radius=2];
\draw[dashed] (0,2) arc [start angle=90,end angle=-90, x radius=1, y radius=2];
\draw[thick,red] (5.25,0) arc [start angle=360,end angle=0, x radius=0.25, y radius=0.5];
\draw (0.14,1.99) to [out=-32,in=180] (5,0.5);
\draw (0.03,-2.02) to [out=31,in=180] (5,-0.5);
\draw (0,-2.1) node [below]{$\bt$};
\draw (5,-0.5) node [below]{$2r$};
\draw (3.5,0) node [left]{$\text{trumpet}$};
\draw (5.4,0) node [right]{$\times ~(-1)^rq^{\hf r(r+1)}$.};
\draw (-1.3,0) node [left]{$\displaystyle ~~=~~\sum_{r\in\mathbb{Z}}$};
\draw (-5.6,0) circle[radius=2];
\draw (-5.6,0) node {disk};
\draw (-5.6,-2.1) node [below]{$\bt$};
\end{tikzpicture}
\label{eq:disk-trumpet}
\end{equation}

More generally, let us consider the operator
\begin{equation}
\begin{aligned}
 H_n(X|q)=\int_0^\pi\frac{d\th}{2\pi}\mu(\th)H_n(\cos\th|q)|\th\ket\bra\th|.
\end{aligned} 
\label{eq:Hn-th-int}
\end{equation}
We can show that this operator can be expanded in terms of the baby universe operators, as follows.
From \eqref{eq:Jacobi} and \eqref{eq:q-H}, we find
\begin{equation}
\begin{aligned}
 \mu(\th)H_n(\cos\th|q)&=
\sum_{r\in\mathbb{Z}}
(-1)^rq^{\hf r(r+1)}\sum_{k=0}^n\frac{(q;q)_n}{(q;q)_k(q;q)_{n-k}}
2\cos[(n+2r-2k)\th]\\
&=\sum_{r\in\mathbb{Z}}
(-1)^rq^{\hf r(r+1)}2\cos[(n+2r)\th]\sum_{k=0}^n\frac{(q;q)_n}{(q;q)_k(q;q)_{n-k}}
(-q^{r+1})^kq^{\hf k(k-1)}\\
&=\sum_{r\in\mathbb{Z}}
(-1)^rq^{\hf r(r+1)}2\cos[(n+2r)\th](q^{r+1};q)_n.
\end{aligned} 
\end{equation}
In the second equality we have shifted $r\to r+k$ and in the third equality
we have used the $q$-binomial formula
\begin{equation}
\begin{aligned}
 \sum_{k=0}^n \frac{(q;q)_n}{(q;q)_k(q;q)_{n-k}}(-a)^k q^{\hf k(k-1)}
=(a;q)_n.
\end{aligned} 
\end{equation}
Finally we arrive at our desired expansion of $H_n(X|q)$
\begin{equation}
\begin{aligned}
 H_n(X|q)=\sum_{r\in\mathbb{Z}}
(-1)^rq^{\hf r(r+1)}(q^{r+1};q)_n\,
\cB_{n+2r}.
\end{aligned} 
\label{eq:Hn-B}
\end{equation}
From the relation
\begin{equation}
\begin{aligned}
 (q^{r+1};q)_n=0,\quad(-1\geq r\geq -n),
\end{aligned} 
\end{equation}
it follows that the size $|n+2r|$ of the baby universe is always larger than $n$
for the operator $\cB_{n+2r}$ with a non-zero
coefficient in \eqref{eq:Hn-B}. 
As a special case, for $n=1$ \eqref{eq:Hn-B} becomes
\begin{equation}
\begin{aligned}
 2X=\sum_{r\in\mathbb{Z}}
(-1)^rq^{\hf r(r+1)}(1-q^{1+r})
\cB_{1+2r}.
\end{aligned} 
\end{equation}
This means that the transfer matrix $T=-E_0X$ is written as a linear
combination of the baby universe operators.
For $n=0$ \eqref{eq:Hn-B} reduces to the expansion of the identity
$\id$ in \eqref{eq:id-B}. 

We can show that the chord number state $|n\ket$ is obtained by 
acting $H_n(X|q)$ on the $0$-chord state $|0\ket$.
Using \eqref{eq:Hn-th-int}, \eqref{eq:n-th} and \eqref{eq:def-id}, we find
\begin{equation}
\begin{aligned}
 \frac{1}{\rt{(q;q)_n}}H_n(X|q)|0\ket&=
\int_0^\pi\frac{d\th}{2\pi}\mu(\th)|\th\ket\bra\th|0\ket \frac{H_n(\cos\th|q)}{\rt{(q;q)_n}}\\
&=\int_0^\pi\frac{d\th}{2\pi}\mu(\th)|\th\ket \bra \th|n\ket \\
&=|n\ket.
\end{aligned} 
\label{eq:Hn-n}
\end{equation}
Using \eqref{eq:Hn-n} and \eqref{eq:Hn-B}, the transition amplitude $\bra n|e^{-\bt T}|0\ket$
between $|0\ket$ and $\bra n|$ is written as
\begin{equation}
\begin{aligned}
 \bra n|e^{-\bt T}|0\ket&=\frac{1}{\rt{(q;q)_n}}\bra 0|H_n(X|q)e^{-\bt T}|0\ket\\
&=\frac{1}{\rt{(q;q)_n}}
\sum_{r\in\mathbb{Z}}
(-1)^rq^{\hf r(r+1)}(q^{r+1};q)_n\bra 0|\cB_{n+2r}
e^{-\bt T}|0\ket\\
&=
\sum_{r\in\mathbb{Z}}
C_{n,r}I_{n+2r}(\bt E_0),
\end{aligned} 
\label{eq:HH}
\end{equation}
where we defined
\begin{equation}
\begin{aligned}
 C_{n,r}=\frac{1}{\rt{(q;q)_n}}(-1)^rq^{\hf r(r+1)}
(q^{r+1};q)_n.
\end{aligned} 
\label{eq:Cnr}
\end{equation}
\eqref{eq:HH} reproduces the result of \cite{Okuyama:2022szh},
in which the amplitude $\bra n|e^{-\bt T}|0\ket$ was interpreted as 
the Hartle-Hawking wavefunction of the bulk theory.
Again, our result \eqref{eq:HH}
implies that the Hartle-Hawking wavefunction 
$\bra n|e^{-\bt T}|0\ket$ is written as a linear combination of trumpets
\begin{equation}
\begin{tikzpicture}[scale=0.75]
\draw (0,2) arc [start angle=90,end angle=270, x radius=1, y radius=2];
\draw[dashed] (0,2) arc [start angle=90,end angle=-90, x radius=1, y radius=2];
\draw[thick,red] (5.25,0) arc [start angle=360,end angle=0, x radius=0.25, y radius=0.5];
\draw (0.14,1.99) to [out=-32,in=180] (5,0.5);
\draw (0.03,-2.02) to [out=31,in=180] (5,-0.5);
\draw (0,-2.1) node [below]{$\bt$};
\draw (5,-0.5) node [below]{$n+2r$};
\draw (3.5,0) node [left]{$\text{trumpet}$};
\draw (5.4,0) node [right]{$\times ~C_{n,r}~~.$};
\draw (-1.3,0) node [left]{$\displaystyle ~~=~~\sum_{r\in\mathbb{Z}}$};
\draw (-5.6,-1.6) node [below]{$\bt$};
\draw (-5.6,0.5) node [above]{$n$};
\draw(-7.6,0.5)--(-3.6,0.5); 
\draw (-7.6,0.5) arc [start angle=-180,end angle=0,radius=2];
\end{tikzpicture}
\label{eq:HH-trumpet}
\end{equation}

Similarly, the bi-local operator 
$\wick{\c\cO_{\lap} e^{-\bt H}\c \cO_{\lap}}$ introduced in 
\cite{Berkooz:2018jqr} can be expanded in terms of the baby universe
operators. Here, the overline of $\wick{\c\cO_{\lap} e^{-\bt H}\c \cO_{\lap}}$
represents the Wick contraction of the random coefficient
$K_{i_1\cdots i_s}$ in $\cO_\lap$.
As shown in \cite{Berkooz:2018jqr},
the bi-local operator 
$\wick{\c\cO_{\lap} e^{-\bt H}\c \cO_{\lap}}$ commutes with $X$
\begin{equation}
\begin{aligned}
 {[}X, \wick{\c\cO_{\lap} e^{-\bt H}\c \cO_{\lap}}{]}=0,
\end{aligned} 
\end{equation}
and hence it becomes diagonal in the $|\th\ket$-basis \cite{Okuyama:2024yya}
\begin{equation}
\begin{aligned}
 \wick{\c\cO_{\lap} e^{-\bt H}\c \cO_{\lap}}=
\int_0^\pi\frac{d\th}{2\pi}\mu(\th)\bra\th|q^{\lap\h{N}}e^{-\bt T}|0\ket
|\th\ket\bra\th|,
\end{aligned} 
\end{equation}
where $\h{N}$ denotes the number operator
\begin{equation}
\begin{aligned}
 \h{N}|n\ket=n|n\ket.
\end{aligned} 
\end{equation}
Inserting the identity operator $\id$ in \eqref{eq:def-id}, we find
\begin{equation}
\begin{aligned}
  \wick{\c\cO_{\lap} e^{-\bt H}\c \cO_{\lap}}&=\sum_{n=0}^\infty q^{\lap n}
\bra n|e^{-\bt T}|0\ket
\int_0^\pi\frac{d\th}{2\pi}\mu(\th) \bra \th|n\ket|\th\ket\bra\th| \\
&=\sum_{n=0}^\infty q^{\lap n}
\bra n|e^{-\bt T}|0\ket \frac{H_n(X|q)}{\rt{(q;q)_n}}.
\end{aligned} 
\end{equation}
From \eqref{eq:Hn-B} and \eqref{eq:HH},
the bi-local operator
$\wick{\c\cO_{\lap} e^{-\bt H}\c \cO_{\lap}}$ is expanded as
\begin{equation}
\begin{aligned}
\wick{\c\cO_{\lap} e^{-\bt H}\c \cO_{\lap}}
=\sum_{n=0}^\infty \sum_{r,s\in\mathbb{Z}} q^{\lap n}
C_{n,r}
C_{n,s}I_{n+2r}(\bt E_0)
\cB_{n+2s}.
\end{aligned} 
\label{eq:bilocal-B}
\end{equation}

\section{Thermofield double state and $\text{ER}=\text{EPR}$}\label{sec:TFD}
In this section, we consider the
thermofield double state $|\text{TFD}\ket$ of DSSYK
and its representation in terms of the baby universe operators.

$|\text{TFD}\ket$ is an element of the doubled Hilbert space
$\cH\otimes\cH$ \footnote{See \cite{Okuyama:2024yya,Okuyama:2024gsn,Lin:2023trc,Xu:2024hoc} for the study of the doubled Hilbert space in DSSYK.}
of the chord Hilbert space $\cH$ in \eqref{eq:H-chord}.
$|\text{TFD}\ket$ is given by a Euclidean time evolution of the maximally
entangled state $|\text{max}\ket$
\begin{equation}
\begin{aligned}
 |\text{TFD}\ket=(e^{-\hf\bt T}\otimes e^{-\hf\bt T})|\text{max}\ket,
\end{aligned} 
\end{equation}
where $|\text{max}\ket$ is given by
\begin{equation}
\begin{aligned}
 |\text{max}\ket=\sum_{n=0}^\infty |n\ket\otimes|n\ket=
\int_0^\pi\frac{d\th}{2\pi}\mu(\th)|\th\ket\otimes|\th\ket.
\end{aligned} 
\label{eq:max}
\end{equation}
Using the last expression of $|\text{max}\ket$ in \eqref{eq:max}, 
$|\text{TFD}\ket$ is written as
\begin{equation}
\begin{aligned}
 |\text{TFD}\ket=
\int_0^\pi\frac{d\th}{2\pi}\mu(\th)e^{-\bt E(\th)}|\th\ket\otimes|\th\ket.
\end{aligned} 
\end{equation}
From the first expression of $|\text{max}\ket$ in \eqref{eq:max},
we find that $|\text{TFD}\ket$ can be written in terms of the
baby universe operators.
Plugging \eqref{eq:Hn-n} and \eqref{eq:Hn-B} into \eqref{eq:max},
we find
\begin{equation}
\begin{aligned}
 |\text{TFD}\ket=\sum_{n=0}^\infty \sum_{r,s\in\mathbb{Z}}
C_{n,r}C_{n,s}(e^{-\hf \bt T}\cB_{n+2r}\otimes e^{-\hf \bt T}\cB_{n+2s})|0\ket\otimes|0\ket.
\end{aligned} 
\label{eq:TFD-BB}
\end{equation}
The operator $e^{-\hf \bt T}\cB_{n+2r}\otimes e^{-\hf \bt T}\cB_{n+2s}$
can be thought of as a creation operator of the double trumpet
\cite{Saad:2018bqo,Saad:2019lba}.
In other words, two sides of the 
thermofield double state in \eqref{eq:TFD-BB}
are connected by a Euclidean wormhole generated by the operator
$e^{-\hf \bt T}\cB_{n+2r}\otimes e^{-\hf \bt T}\cB_{n+2s}$.
This is schematically depicted as
\begin{equation}
\begin{tikzpicture}[scale=1.1]
\draw (-3,-1.6) node [below]{$\bt$};
\draw (-3,0) node [below]{TFD};
\draw(-5,0.5)--(-1,0.5); 
\draw (-5,0.5) arc [start angle=-180,end angle=0,radius=2];
\draw (0,0) node [below]{$=$};
\draw (1,0.5) arc [start angle=-180,end angle=0,radius=2];
\draw(1,0.5)--(5,0.5); 
\draw[dashed](3,0.5)--(3,-0.3); 
\draw[dashed](3,-0.8)--(3,-1.5); 
\draw (1.8,-0.4) to  [out=-30, in=-150] (4.2,-0.4);
\draw (2,0) to  [out=-30, in=-150] (4,0);
\draw[red, thick] (1.9,-0.2) circle [x radius=0.15,y radius=0.22,rotate=-30];
\draw[red, thick] (4.1,-0.2) circle [x radius=0.15,y radius=0.22,rotate=30];
\draw (1.5,-1) node [below]{$\hf\bt$};
\draw (4.5,-1) node [below]{$\hf\bt$};
\end{tikzpicture}
\label{eq:TFD-pic}
\end{equation}
where the red circles on the right hand side represent the baby universes
created by the operator $\cB_{n+2r}\otimes \cB_{n+2s}$
and the sum over $n$ in \eqref{eq:TFD-BB}
gives rise to a double trumpet, or a Euclidean wormhole.

Our expression \eqref{eq:TFD-BB}
can be thought of a concrete realization of the idea of $\text{ER}=\text{EPR}$
\cite{VanRaamsdonk:2010pw,Maldacena:2013xja},
represented by a geometric connection by a wormhole in \eqref{eq:TFD-pic}.
We also notice that the bi-local operator in \eqref{eq:bilocal-B}
has a similar structure
as \eqref{eq:TFD-BB}, where one of the baby universe operator
$\cB_{n+2r}$ is replaced by the trumpet $I_{n+2r}(\bt E_0)$.

\section{Comments on JT gravity}\label{sec:JT}
As discussed in \cite{Berkooz:2018jqr,Lin:2022rbf},
JT gravity is obtained from DSSYK by the triple scaling limit
\begin{equation}
\begin{aligned}
 \quad\la\to0,\quad n\to\infty,\quad\th\to0,
\end{aligned} 
\label{eq:triple}
\end{equation}
with the following $\ell$ and $k$ held fixed
\begin{equation}
\begin{aligned}
 \ell=\la n+2\log\la,\quad k=\frac{\th}{\la}.
\end{aligned} 
\label{eq:ell}
\end{equation}
This limit \eqref{eq:triple} amounts to zooming in on the edge $E(\th)=-E_0$
of the spectrum $E(\th)$ in \eqref{eq:E-th}.
In other words, JT gravity is obtained as a low energy limit of DSSYK.
One can show that
 $\bra\th|n\ket$ in \eqref{eq:n-th}
and $\mu(\th)$ in \eqref{eq:mu-def} reduce to the wavefunction $\bra k|\ell\ket$
and
the spectral density $\rho(k)$ of JT gravity
in this limit \eqref{eq:triple}
\begin{equation}
\begin{aligned}
 \bra\th|n\ket&~\to ~\bra k|\ell\ket=K_{2\ri k}(2e^{-\hf\ell}),\\
\mu(\th)&~\to ~\rho(k)=\frac{2k}{\pi}\sinh(2\pi k),
\end{aligned} 
\label{eq:rho-JT}
\end{equation}
where $K_\nu(z)$ denotes the modified Bessel function
of the second kind.
Similarly, $\cB_a$ in \eqref{eq:Ba-def}
reduces to the baby universe operator $\cO_b$ in JT gravity
\cite{Penington:2023dql}
\begin{equation}
\begin{aligned}
 \cB_a~\to ~\cO_b=\int_0^\infty\frac{dk}{2\pi}\,2\cos(bk)|k\ket\bra k|,
\end{aligned} 
\end{equation}
where $a$ and $b$ are related by
\begin{equation}
\begin{aligned}
 b=\la a.
\end{aligned} 
\label{eq:b-a}
\end{equation}

The relation \eqref{eq:cos/mu} in DSSYK has an analogue in JT gravity
\footnote{This relation \eqref{eq:K-int}
was utilized in the study of FZZT branes in minimal string theory \cite{Martinec:2003ka}.}
\begin{equation}
\begin{aligned}
 \frac{2\cos(bk)}{\rho(k)}=\int_{-\infty}^\infty d\ell\, 
K_{2\ri k}(2e^{-\hf\ell}) e^{-2e^{-\hf\ell}\cosh\frac{b}{2}}.
\end{aligned} 
\label{eq:K-int}
\end{equation}
Thus $\cO_b$ in JT gravity can be written as a combination
of the Hamiltonian
\begin{equation}
\begin{aligned}
 H=\int_0^\infty\frac{dk}{2\pi}\rho(k)E(k)|k\ket\bra k|,\quad E(k)=k^2,
\end{aligned} 
\end{equation}
in a similar manner as \eqref{eq:Ba-X}.

However, there is no counterpart of \eqref{eq:Jacobi}
in JT gravity, 
since $\rho(k)$ in \eqref{eq:rho-JT} exponentially grows at large $k$
and hence it is not Fourier-transformable, i.e.,
$\til{\rho}(b)$ satisfying the following equation does not exist
\begin{equation}
\begin{aligned}
 \rho(k)=\int_0^\infty db\,\til{\rho}(b)\cos(bk).
\end{aligned} 
\end{equation}
Therefore, there is no natural way to connect
the disk partition function and the trumpet in JT gravity
as in \eqref{eq:disk} for DSSYK.
This difference comes from the fact that 
the spectrum of DSSYK is bounded $|E|\leq E_0$,
while that of JT gravity is unbounded
and $\rho(k)$ grows exponentially at large $k=\rt{E}$. 
We should stress that the expansion of the identity $\id$
in terms of $\cB_a$ in \eqref{eq:id-B}
is a special feature of DSSYK, which has no counterpart in JT gravity.
 
\section{Conclusions and outlook}\label{sec:conclusion}
In this paper, we have studied the baby universe operator 
$\cB_a$ in DSSYK.
We found that $\cB_a$ is written in terms of the transfer matrix
$T=-E_0X$, and vice versa.
In particular, the identity operator $\id$ and $H_n(X|q)$
are expanded as a linear combination
of $\cB_a$ (see \eqref{eq:id-B} and \eqref{eq:Hn-B}). 
This implies that the disk partition function $\bra 0|e^{-\bt T}|0\ket$
and the Hartle-Hawking wavefunction $\bra n|e^{-\bt T}|0\ket$
are expanded in terms of trumpets (see \eqref{eq:disk} and \eqref{eq:HH}).
We also found that the thermofield double state of DSSYK
in \eqref{eq:TFD-BB} is generated by the 
double trumpet operator $\cB_{n+2r}\otimes \cB_{n+2s}$.
This can be thought of as a concrete realization of the idea of 
$\text{ER}=\text{EPR}$ \cite{VanRaamsdonk:2010pw,Maldacena:2013xja}.

There are many interesting open questions.
As argued in \cite{Penington:2023dql}, the baby universe operator
in JT gravity is ill-defined in the presence of bulk matter fields
due to the UV divergence coming from the matter loop running around the
neck of a wormhole.
We expect that the baby universe operator
in DSSYK has a much better UV behavior since
the bulk geodesic length of DSSYK is discretized in units of $\la$
and hence $\la$ serves as a natural UV cut-off 
(see \eqref{eq:ell} and \eqref{eq:b-a}).
It would be interesting to study the loop correction to
the baby universe operator in DSSYK along the lines of 
\cite{Jafferis:2022wez,Okuyama:2023aup,Okuyama:2023yat}.

In \cite{Stanford:2022fdt},
the firewall paradox of black hole was studied by
computing the baby universe emission
amplitude in JT gravity
(see also \cite{Zolfi:2024ldx,Blommaert:2024ftn,Iliesiu:2024cnh}).
It was argued in \cite{Stanford:2022fdt} that 
there appears a firewall by the ``wormhole shortening effect''
due to the baby universe emission \cite{Saad:2019pqd}.
It would be interesting to
repeat a similar computation of the baby universe emission
in DSSYK and see if the firewall
appears in DSSYK as well.
Work in this direction is in progress and 
will be reported elsewhere
\cite{Miyaji}.

\acknowledgments
The author would like to thank Masamichi Miyaji, Soichiro Mori and Masaki Shigemori 
for discussions. 
This work was supported
in part by JSPS Grant-in-Aid for Transformative Research Areas (A) 
``Extreme Universe'' 21H05187 and JSPS KAKENHI 22K03594.

\bibliography{paper}
\bibliographystyle{utphys}

\end{document}